\documentclass{acm_proc_article-sp}
\usepackage{hyperref}
\usepackage{amsmath}

\begin{document}

\title{Storing complex data sharing policies with the Min Mask Sketch}

\numberofauthors{2} 
\author{
\alignauthor
Stephen Smart\\
       \affaddr{University of Oklahoma}\\
       \affaddr{School of Computer Science}\\
       \email{smart@ou.edu}
\alignauthor
Christan Grant\\
       \affaddr{University of Oklahoma}\\
       \affaddr{School of Computer Science}\\
       \email{cgrant@ou.edu}
}

\date{20 April 2013}

\maketitle
\begin{abstract}
More data is currently being collected and shared by software applications than ever before.
In many cases, the user is asked if either all or none of their data can be shared.
We hypothesize that in some cases, users would like to share data in more complex ways.
In order to implement the sharing of data using more complicated privacy preferences, complex data sharing policies must be used.
These complex sharing policies require more space to store than a simple ``all or nothing'' approach to data sharing.
In this paper, we present a new probabilistic data structure, called the Min Mask Sketch, to efficiently store these complex data sharing policies.
We describe an implementation for the Min Mask Sketch in PostgreSQL and analyze the practicality and feasibility of using a probabilistic data structure for storing complex data sharing policies.
\end{abstract}

\section{Introduction}

The storage and management of large data sets is becoming increasingly common.
Many applications are continuously recording data about its users and sharing this data to other entities.
This leads to data privacy issues and as more data driven applications are coming into existence, these privacy issues are becoming more complex.
One approach to handling data privacy when it comes to managing and sharing user data is a simple ``all or nothing'' approach.
In other words, all of the data can be shared or all of it is restricted.
This approach works for many applications, but what if the user would like to share a portion of the data being recorded and hide the rest?
What if the user would like to share her data in a more complex manner such as dependent on time, location, or a combination of several conditions?
These complex policies for data sharing are becoming more practical with the development of more data driven applications and the growth of the underlying network in which these applications communicate, i.e. the Internet of Things.

Complex data sharing policies such as those mentioned above are difficult to implement in a modern database management system.
In addition to the overhead added to the development life cycle, complex sharing policies also require more space. 
Instead of a simple Boolean value representing the ``all or nothing'' approach to data privacy described above, more bits are needed to represent these policies and every policy could potentially be unique to a single data point in the data set.
Many of these data driven applications that are recording user's personal data exist in the mobile application domain, therefore space is an important consideration.

In this paper, we will describe one approach to improve the space efficiency of storing complex data sharing policies. 
This approach involves the development and use of a novel probabilistic data structure, that we will call the Min Mask Sketch, to store complex sharing policies in a small amount of space. 
As with most probabilistic data structures, a small amount of accuracy will be sacrificed in exchange for an increase in space efficiency. 
One of the most popular probabilistic data structures is the Bloom Filter~\cite{bloom1970space}. 
The goal of the Bloom Filter is to determine if any given item is a member of a large data set without having to store the entire data set in memory. 
Over the years, many new probabilistic data structures have been developed that implement a similar approach used in the Bloom Filter but include strategic modifications to answer a different question about the original data. 
One such data structure is the Count Min Sketch, which not only answers the question of set membership, but additionally can determine the frequency at which a given item exists in the data set. 
The Min Mask Sketch is a modified version of the Count Min Sketch~\cite{cormode2005improved} that can be used to determine a given item's privacy policy.

The remaining sections of this paper will be organized as follows:
\begin{itemize}
    \item Section~\ref{sec:complexsharingpolicies} discusses the idea of complex data sharing policies in more detail and describes some of background work that sparked many of the ideas introduced in this paper.
    \item Section~\ref{sec:example} introduces an example application and relational schema to illustrate one potential practical application of the Min Mask Sketch approach to storing complex data sharing policies.
    \item Section~\ref{sec:minmasksketch} explains the Min Mask Sketch data structure in detail.
    \item Section~\ref{sec:implementation} describes our implementation of the Min Mask Sketch data structure in PostgreSQL 9.6
    \item Section~\ref{sec:analysis} analyzes the feasibility and practical applications of the Min Mask Sketch approach and compares this approach with some alternative methods.
    \item Section~\ref{sec:conclusion} summarizes the approach and provides concluding remarks.
\end{itemize}

\section{Complex Sharing Policies}
\label{sec:complexsharingpolicies}
We define complex data sharing as the sharing of data that requires fine grained access control. 
In other words, each individual data point could be restricted based on a different set of conditions. 
When data is shared in this way, the standard approach to data privacy does not work. 
More sophisticated approaches that use complex sharing policies must be used.

In Appendix~\ref{sec:complexpolicyexamples} we give examples of sharing policies that we label as complex.
We show how complex policies can be created from a combination of fundamental policy demands.
We walk through pictorial representations of both the fundamental sharing policies and more complex policies.

Work has been done in the database community to develop methods for implementing data sharing policies within Hippocratic database systems~\cite{lefevre2004limiting}.
Language constructs have been created to define these fine grained access control policies with minimal complexity~\cite{agrawal2005extending}.
One goal for minimizing the complexity of these policy representations is to reduce the storage overhead on a database management system that implements fine grained access control.
In this paper, we introduce a new method for storing policy meta data that aims to further reduce the cost of storage.

\section{Example}
\label{sec:example}

Consider a new mobile application, Health Tracker Pro, that uses a device comparable to a Fitbit to record a user's health data. 
The purpose of this application is to not only help users monitor their personal health, but also give them the ability to share their personal health data with their doctor. 
Doctor's would use the Health Tracker Pro Dashboard application to view health data shared by each of their patients.

The primary data recorded by Health Tracker Pro is stored in a single table. 
This table has the following schema (using PostgreSQL data types):
\begin{verbatim}
    health_data(
        time         timestamp   primary_key
        heart_rate   smallint    not null
        blood_sugar  smallint    not null
        body_temp    real        not null
    )
\end{verbatim}

Health Tracker Pro uses a sampling rate of 20 times per minute, or more precisely, once every three seconds. 
An example subset of this data can be seen in Figure 1 for an example user, Bob.
\begin{figure}
  \includegraphics[width=\linewidth]{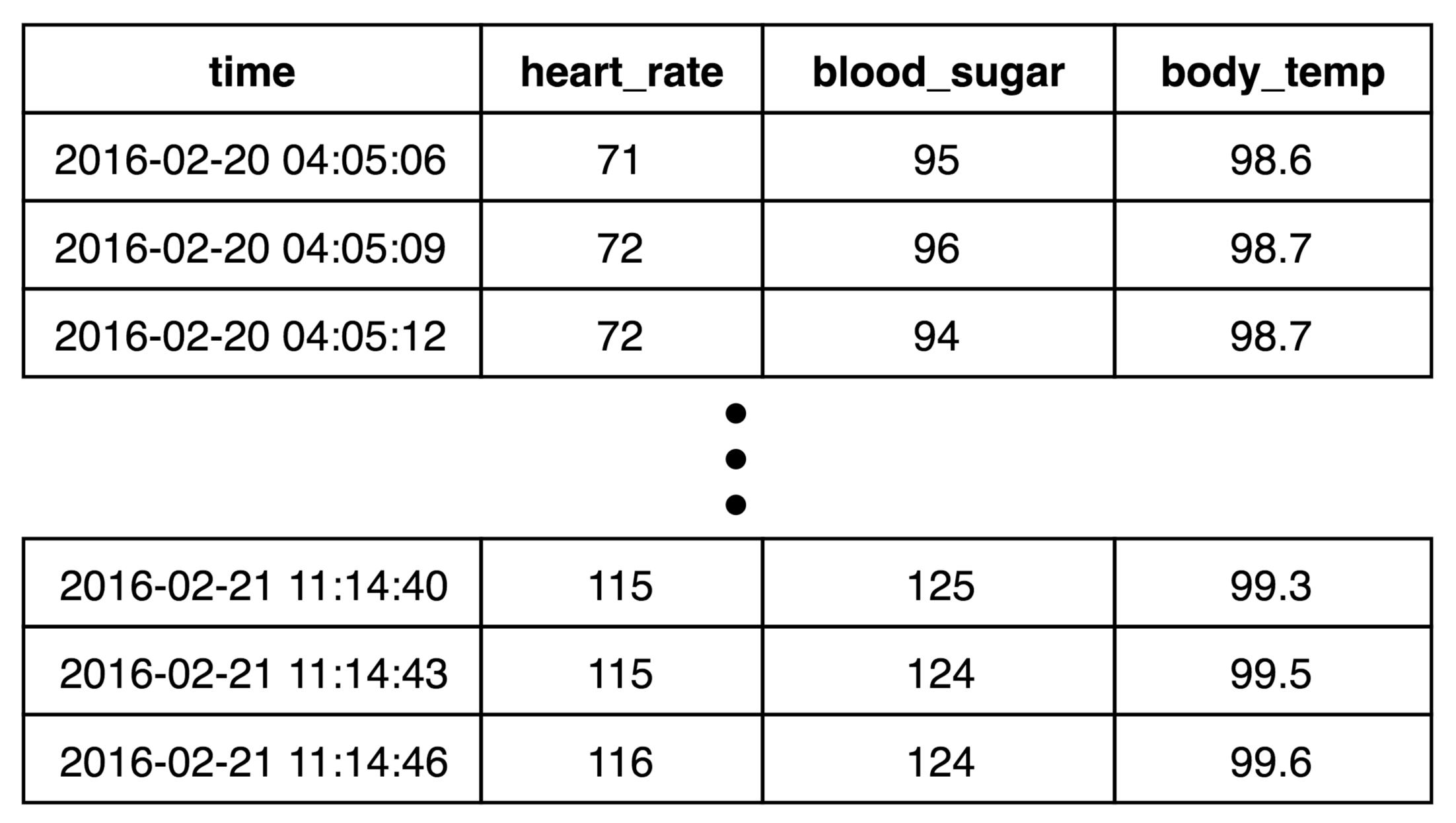}
  \caption{Example table containing Bob's personal health data recorded by Health Tracker Pro.}
  \label{fig:example}
\end{figure}

Bob would like to share some of his health data with his doctor.
However, he does not want to share all of the data recorded by Health Tracker Pro. 
Bob only wishes to share his data at certain times during the day.
For example, Bob would like to share his heart rate and body temperature data while exercising, and blood sugar data while sleeping and after eating a meal. 
Additionally, if Bob's heart rate is recorded to be outside of a selected window, he would like his doctor to be notified. 
At all other times of the day, Bob would like his personal health data to remain private.

The simplest approach to storing these complex privacy policies would be by adding three new attributes to the health\textunderscore data table as shown below:
\begin{verbatim}
    health_data(
        time         timestamp    pimary_key
        heart_rate   smallint     not null
        blood_sugar  smallint     not null
        body_temp    real         not null
        hr_private   boolean      not null
        bs_private   boolean      not null
        bt_private   boolean      not null
    )
\end{verbatim}

These attributes are simple Boolean values determining whether the corresponding attribute for a row is private. 
Each row may represent a unique policy, so the policy data will be stored in the same table alongside the primary health data.
In the following sections, we will describe the Min Mask Sketch data structure and implementation that can store these privacy policies without using near as much space. 
We will then discuss the pros and cons of this data structure when compared to the simple method described here as well as briefly mention another alternative approach to storing these complex sharing policies.

\section{Min Mask sketch}
\label{sec:minmasksketch}

\begin{figure}
  \includegraphics[width=\linewidth]{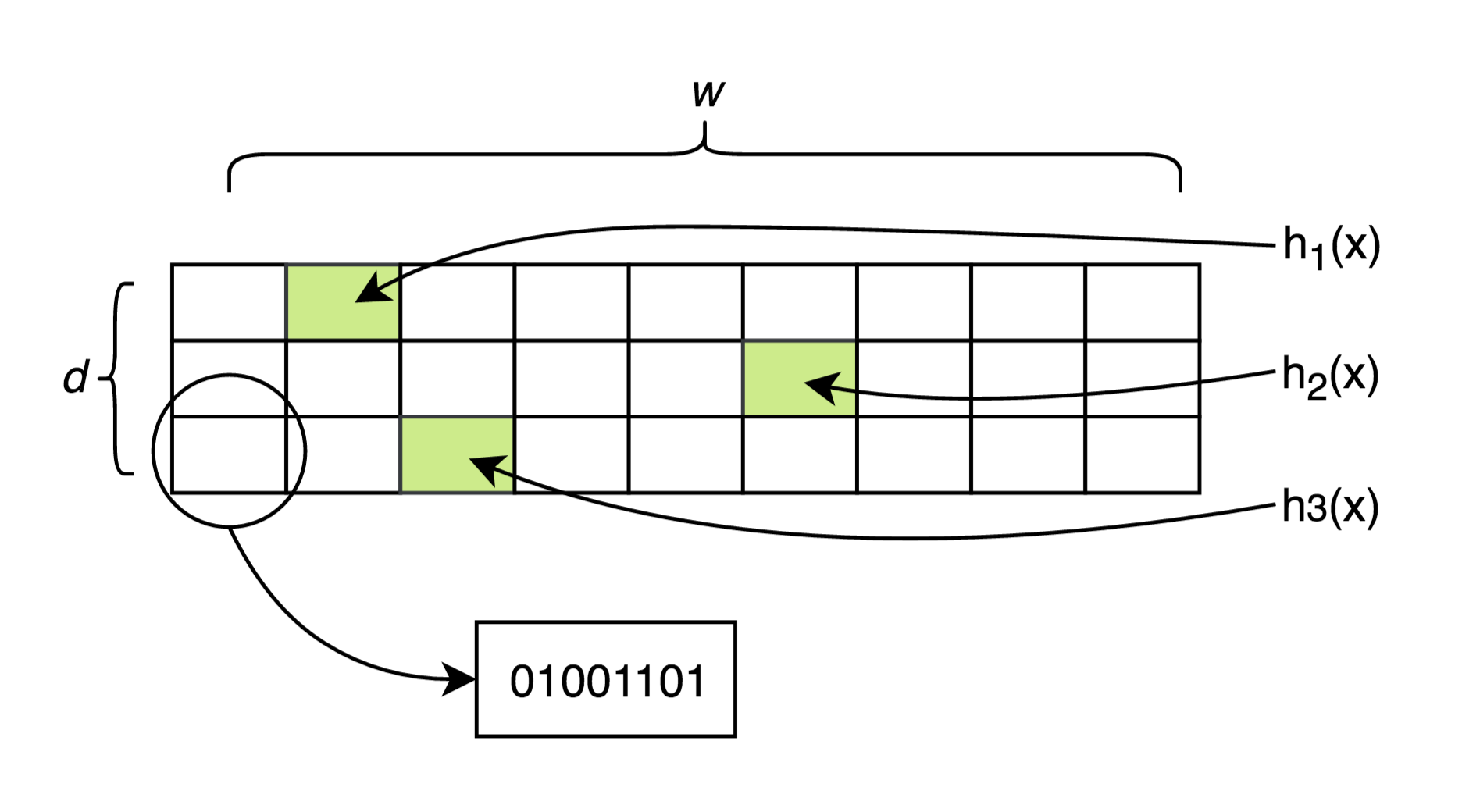}
  \caption{The Min Mask Sketch}
  \label{fig:minmasksketch}
\end{figure}

The Min Mask Sketch is a modified version of the Count Min Sketch~\cite{cormode2005improved}.
The Min Mask Sketch is stored as a two-dimensional array of unsigned integers and uses a collection of hash functions. 
The purpose of the Min Mask Sketch is to efficiently store policies associated with items in a large data set. 
When the sketch is first created, all elements in the two-dimensional array are initialized to zero. 
When a new item's policy is inserted into the sketch, the item is hashed by $d$ different hash functions, where $d$ is the number of rows in the two-dimensional array. 
These hash functions return a uniformly random value between 0 and $w$-1, where $w$ is the number of columns in the two-dimensional array. 
The policy for the given data item is then inserted into a cell contained in each row of the two-dimensional array at the particular index calculated by the corresponding hash function. This process is illustrated in Figure~\ref{fig:minmasksketch}. 

At time of insertion, the policy should be in the form of an unsigned integer. 
We will refer to this unsigned integer as the bitmask for that policy. 
The bitmask approach is very simple. 
Each bit position in the bitmask corresponds to a possible condition in the complex sharing policy associated with a data item. 
If the bit at a particular position is 1, then the condition corresponding to that bit position is \textit{active} for that item and should be applied when the data is shared. 
To insert the policy value into a cell, a bitwise OR operation is performed with the existing bitmask in the cell and the new bitmask to be inserted. 
This is done to avoid overwriting existing bitmasks in the sketch when a hash collision occurs or when performing an update to an existing item.  
Inserting a new item into the Min Mask Sketch is a straightforward process.
Updating an existing item presents new problems which will be discussed in Section~\ref{sec:analysis}.

Using the running example from Section~\ref{sec:example}, suppose the least significant bit position in the bitmask corresponds to making Bob's body temperature information private, the next highest order bit position corresponds to making Bob's blood sugar information private, and the next highest order bit position corresponds to making Bob's heart rate information private. 
During exercise, Bob would like to share his heart rate and body temperature information with his doctor. 
The corresponding sharing policy for each data item recorded during his exercise session would be ``010''. 
Meaning that his heart rate and body temperature are freely being shared, but his blood sugar has an active ``private'' condition associated with it that should be applied during the sharing process. 
Since the time attribute is the primary key for the health\textunderscore data table in the example given in Section~\ref{sec:example}, the time value for each data item would be the argument passed into each hash function.

When retrieving a policy bitmask for a given item, the primary key for that item is hashed by all $d$ hash functions to get the indexes that should be checked for each row in the two-dimensional array.
The bitmask values at each index are passed into a function that determines the number of 1's, or ``active policies'', contained in each bitmask and returns the bitmask with the minimum number of active policies as the result. 
Due to hash collisions, this bitmask is only an estimate for the actual bitmask associated with the given data item. 
This estimate is bounded by the following equation with probability $c$:
\begin{equation}
\label{equ:estimatebound}
a \leq \hat{a} \leq a + \epsilon n
\end{equation}

Where $a$ is the actual policy for a given item, $\hat{a}$ is the estimated policy, $\epsilon$ is the error bound factor and n is the number of insertions into the Min Mask Sketch. 
This equation has been formally proven by Cormode and Muthukrishnan in the original paper describing the Count Min Sketch~\cite{cormode2005improved}. 
Since this equation is known to be true, the estimate for the policy associated with a given item in the sketch will always be either the correct policy or contain a small amount of extra 1's in the policy.
The design of the Min Mask Sketch was done in such a way to error on the side of caution. 
Instead of sharing data that the user wanted to hide, inaccurate estimates will hide data that could have been shared.

$c$, the confidence interval for the error bound, and $\epsilon$,the error bound factor, can be chosen at creation time to fit the sketch with the needs of the particular application. 
The smaller the error bound factor, and the greater the confidence interval, the more space is needed for the Min Mask Sketch to deliver these guarantees. 
This is because the width and depth of the two-dimensional array used for the Min Mask Sketch are determined based on these two parameters:
\begin{equation}
\label{equ:sketchwidth}
w = \lceil \frac{e}{\epsilon} \rceil
\end{equation}
\begin{equation}
\label{equ:sketchdepth}
d = \ln(\frac{1}{1 - c})
\end{equation}

Cormode and Muthukrishnan did extensive theoretical analysis, proving that when the sketch is sized in this manner and $d$ hash functions are used, \eqref{equ:estimatebound} holds true. 
The size of the resulting Min Mask Sketch does not grow as a function of the data set, it is completely fixed based on the tuning of $c$ and $\epsilon$. 
However, $\epsilon$ should be chosen with the number of insertions kept in mind. 
Since the upper bound for the estimation error is a result of multiplying the number of insertions by the error bound factor $\epsilon$, a value for $\epsilon$ can be chosen to tune this upper bound to a precise point based on the estimated number of insertions that will occur in the Min Mask Sketch. 
If $\epsilon$ is chosen to be too large compared to the number of insertions, the upper bound for the error in estimating a policy will grow significantly and the Min Mask Sketch will no longer be useful.

\section{Implementation}
\label{sec:implementation}

Our implementation of the Min Mask Sketch was done by creating an extension for PostgreSQL version 9.6~\footnote{The implementation is available on GitHub~\url{https://github.com/oudalab/mms}.}. 
The extension was written in C and contains four major components:
\begin{enumerate}
    \item Definition of the Min Mask Sketch data type.
    \item Functions to create a new Min Mask Sketch object.
    \item Functions to add an item into the Min Mask Sketch.
    \item Functions to retrieve the bitmask for a given item in the Min Mask Sketch.
\end{enumerate}

The first component of the extension was implemented by creating a simple C structure containing three fields: two integer variables to hold the sketch depth and sketch width, and an array of integers to represent the sketch itself. 
This C structure is then mapped to a PostgreSQL data type called ``mms'' that can be attributed to a column in a CREATE TABLE statement. 
For example, to create a table containing a column with the Min Mask Sketch data type, the following SQL can be executed:
\begin{verbatim}
              CREATE TABLE example (
                  my_sketch mms
              );
\end{verbatim}

Creating this table does not automatically instantiate a new Min Mask Sketch object. 
This is where the second component of the extension is required. In order to instantiate a new Min Mask Sketch object, we created a user-facing function called ``mms'' that accepts two parameters. 
These parameters are floating point numbers corresponding to the error bound and confidence interval for the sketch. 
These are optional parameters with default values of 0.001 and 0.99 respectively. 
The error bound and confidence interval are then used to determine the sketch depth and sketch width. 
This process is discussed in detail in Section~\ref{sec:minmasksketch}. 
The required amount of memory for the sketch array is then allocated and each value in the sketch is initialized to 0. 
The new Min Mask Sketch object is then returned. 
In order to insert a new Min Mask Sketch object into the example table created above, the following SQL code can be executed:
\begin{verbatim}
        INSERT INTO example VALUES(mms());
\end{verbatim}

The third component of the extension handles adding new items into the sketch, and was implemented by creating a user-facing function called ``mms\textunderscore add''. 
This function takes three parameters: the sketch to which the new item should be added, the new item itself, and a bitmask to identify the policy that should be applied to the item when being shared (as described in Section~\ref{sec:minmasksketch}). 
The new item is first hashed, using MurmurHash3, to $d$ different locations in the sketch, where $d$ corresponds to the sketch depth calculated at creation time. 
The values computed by the hash functions correspond to indexes in the sketch array. 
A bitwise OR operation is then performed between the existing value in the sketch at each index and the new bitmask value given as the third function argument. 
This successfully adds or updates the item accordingly. 
An example SQL statement to add a new item into a Min Mask Sketch object is given below:
\begin{verbatim}
      UPDATE example SET my_sketch = 
          mms_add(my_sketch, "abc"::text, 6);
\end{verbatim}

Note in the example above that the new item is of type ``text'', but any data type is supported, and the integer ``6'' corresponds to the binary representation ``110'', meaning two conditions are active for that item. 
It is possible to update an item to use a new policy, however, the update must only result in bits changing from 0 to 1, not the reverse. 
In other words, new conditions for a specific row can be set to active but new existing active conditions cannot be deactivated. 
This limitation is discussed in more detail in Section~\ref{sec:analysis}.

The fourth component of the extension was implemented by creating a user-facing function to retrieve the bitmask associated with a given item in the data set. 
This function is called ``mms\textunderscore get\textunderscore mask'' and takes two parameters. 
The first parameter is the sketch in which the item is stored, and the second parameter is the item in question. 
This function hashes the given item to obtain the $d$ different hash values corresponding to the indexes in the sketch that must be checked. 
The bitmasks at each index are retrieved from the sketch and the minimum mask value is calculated (as described in Section~\ref{sec:minmasksketch}). 
This minimum mask value is then returned to the user as the policy for that data item. 
An example SQL statement to retrieve the bitmask value associated with an item in a Min Mask Sketch is as follows:
\begin{verbatim}
    SELECT mms_get_mask(my_sketch, "abc"::text)
        FROM example;
\end{verbatim}

This query returns the bitmask value associated with the item ``abc'' which can then be used to determine the policy that should be applied to the row identified by ``abc''. 
If the item does not exist in the data set, this function will return 0.

\section{Analysis}
\label{sec:analysis}

After analyzing this approach to store complex sharing policies efficiently, several issues have arisen that will be discussed in this section. 
The first issue is the limitation that the Min Mask Sketch has in regards to handling updates and deletions. 
The Min Mask Sketch approach only succeeds in handling updates that result in changing bits from a 0 to a 1 in the bitmask and not the reverse. 
The sketch cannot handle deletions at all. 
This is due to the fact that when retrieving a policy for a given data item, the policy corresponding to the minimum number of 1's is chosen. 
Therefore, if a policy was updated from ``111'' to ``001'', and one of the indexes calculated by the hash functions collided with a second item in the sketch, this could potentially ruin the accuracy of that second item. 
If the second item involved in the hash collision had a policy of ``101'', one of its $d$ bitmasks would be changed to ``001'' by the update to the first policy, resulting in a new minimum bitmask for the second item that is inaccurate. 
Deleting an item is a problem for the exact same reason, because it results in bitmasks going from a higher number of 1's to a lower number of 1's. The Count Min Sketch and other modifications to the Bloom Filter such as the Counting Bloom Filter are able to handle updates and deletions using increments and decrements~\cite{fan2000summary}. The Min Mask Sketch, however, does not inherit this functionality because performing a bitwise logical OR operation between two integers is not the same as incrementing and thus information can be lost when a hash collision occurs.
One approach to solving this problem would be by choosing the average bitmask value instead of the minimum, however this would result in a looser error bound and thus more inaccuracies.

The second issue with this approach to storing complex policies is the fact that the simplest way of storing the policy does not add a large amount of overhead, so the introduction of inaccuracies when applying sharing policies may not be worth it.
When considering the example from Section~\ref{sec:example}, one entry in the health\textunderscore data table can be stored in 16 bytes (excluding the three additional columns added to store policies), while the policy data can be stored in 3 bytes using the simple approach of adding three extra columns. 
With an overly generous assumption that the Min Mask Sketch approach could store all of the policies in a negligible amount of space, this would result in an 18.75\% space efficiency increase.
This efficiency is reasonable, but it should be noted that this percentage value is an upper bound to the space efficiency increase for this example and would only shrink as more health data was recorded by the Health Tracker Pro app. 
Since the Min Mask Sketch approach brings potential inaccuracies in the estimation of policies, it should result in a space efficiency increase large enough to warrant those inaccuracies. 
If a more complex policy were used that would require even more data to represent it than the primary data itself, the Min Mask Sketch approach to storing these policies would become much more feasible, but practical examples that involve such large complex policies are scarce.

\begin{figure}
  \includegraphics[width=\linewidth]{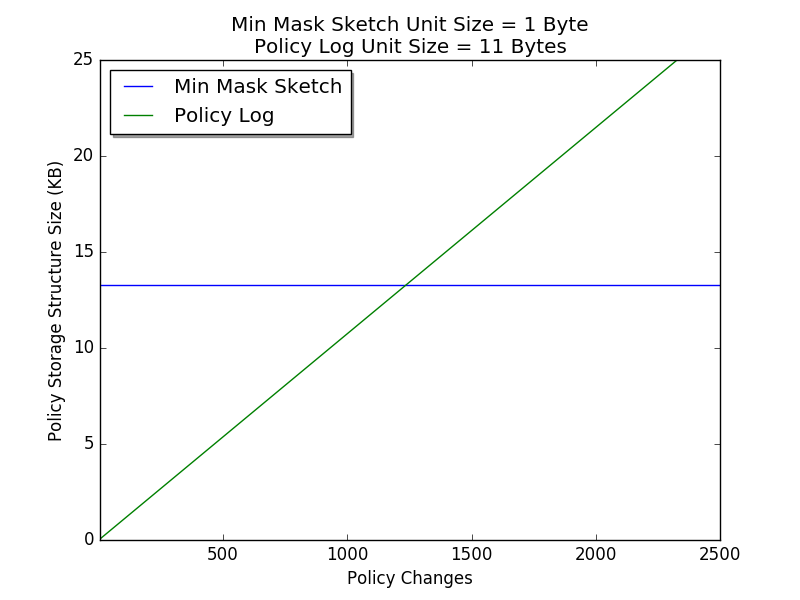}
  \caption{Comparing the space efficiency of the Min Mask Sketch vs the log-based storage approach.}
  \label{fig:comparing}
\end{figure}

The final issue that will be analyzed here is related to the frequency of policy changes within a large data set and their role in the feasibility of the Min Mask Sketch approach. Consider the running example from Section~\ref{sec:example}. 
Based on Bob's wishes, the policies associated with each row would only change a few times per day. 
This means that most rows in the health\textunderscore data table will contain the same complex policy. 
When there are very few policy changes, an alternative method for storing these policies could be used that is based on storing a policy for a range of items in the table. 
We will call this approach the log-based approach. 
For example, if Bob's sharing policy only changes 6 times per day, 6 entries in a log table could be inserted, where each entry contained a timestamp, and a boolean value for each condition. 
The time between each entry in the log table would be the range for those policies to be applied. 
When determining the policy for a given data item, the log table could be referenced and the range of times given by the different timestamps would determine which policy should be applied to that item. 
This approach is much more efficient than other approaches when the frequency of policy changes is low.

The size of the log table grows as a function of the number of policy changes within a data set.
Figure~\ref{fig:comparing} shows the storage space used for the Min Mask Sketch compared to the space used for the log-based approach for a variety of policy changes based on the running example from Section~\ref{sec:example}. 
In this graph, the default error bound factor and confidence interval were used (0.001 and 0.99 respectively). 
As one can clearly see, the log-based approach outperforms the Min Mask Sketch approach for the Health Tracker Pro example until roughly 1250 policy changes occur in the data set. 
Also, the Min Mask Sketch approach introduces inaccuracies due to the probabilistic behavior of the data structure, therefore, the Min Mask Sketch storage approach would need to significantly outperform the log-based approach for it to be a practical choice.

\section{Conclusion}
\label{sec:conclusion}

Complex data sharing policies are becoming increasingly common as more applications are recording data and sharing it across a large network of devices and people. 
We have presented the Min Mask Sketch approach to efficiently store these policies. 
We have also described our implementation for the Min Mask Sketch within the PostgreSQL 9.6 database management system. 
After a detailed analysis of some of the key factors involved in storing complex sharing policies, we have seen that there are several issues with this probabilistic approach to storing complex sharing policies. 
We have discussed these issues in detail in order to understand the fundamental questions that need to be answered when developing solutions for storing complex sharing policies. 
Some of the problems discussed in the analysis section can be solved through future work and design changes, while other problems require a better understanding of how data might be shared in the future in order to solve.

\bibliographystyle{abbrv}
\bibliography{sample}

\appendix

\section{Sharing Policy Examples}
\label{sec:complexpolicyexamples}

In this section, we describe the possible complex sharing policies.
The sharing policies describe the different ways users can control the access to their data.
In the examples, we describe the basic sharing policies and describe how combining sets of policies can create more complex policies.

\subsection{Fundamental Policies}

\begin{figure}[h]
  \includegraphics[width=\linewidth]{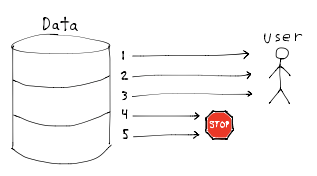}
  \caption{Sharing a limited number of records}
  \label{fig:policy1}
\end{figure}

\paragraph{Sharing a Limited Number of Records}
This policy shown in Figure~\ref{fig:policy1} describes sharing a set number of records from a data set and restricting the rest. 
For example, an owner of a data set would like to share a maximum of 100 records.

\begin{figure}[h]
  \includegraphics[width=\linewidth]{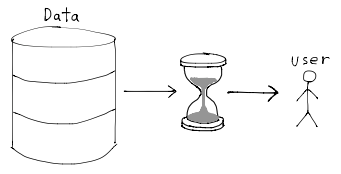}
  \caption{Sharing Data for a Limited Time Period}
  \label{fig:policy2}
\end{figure}

\paragraph{Sharing Data for a Limited Time Period}
This policy shown in Figure~\ref{fig:policy2} describes sharing records from a data set for a limited time period. 
For example, an owner of a data set would like to share data for 24 hours after which the shared data becomes private.

\begin{figure}[h]
  \includegraphics[width=\linewidth]{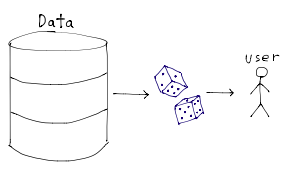}
  \caption{Sharing a Random Sample of Records}
  \label{fig:policy3}
\end{figure}

\paragraph{Sharing a Random Sample of Records}
This policy shown in Figure~\ref{fig:policy3} describes sharing a randomly selected subset of data. 
For example, an owner of a data set may not want to share all of his/her data, but instead chooses to share 25 randomly selected records from the data set.

\begin{figure}[h]
  \includegraphics[width=\linewidth]{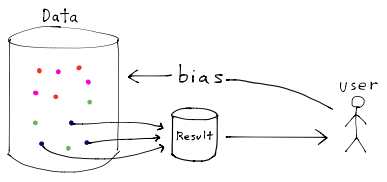}
  \caption{Sharing a Biased Sample of Records}
  \label{fig:policy4}
\end{figure}

\paragraph{Sharing a Biased Sample of Records}
This policy shown in Figure~\ref{fig:policy4} describes sharing a subset of data according to some bias. 
For example, a user located in New York City queries a weather data set, but only the records relevant to New York City and the surrounding area are shared.

\begin{figure}[h]
  \includegraphics[width=\linewidth]{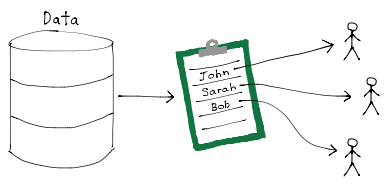}
  \caption{Sharing Data with a Set of Other Users}
  \label{fig:policy5}
\end{figure}

\paragraph{Sharing Data with a Set of Other Users}
This policy shown in Figure~\ref{fig:policy5} describes sharing data to a specific set of users. 
For example, a Facebook user would only like his/her data to be shared with users on his/her friends list. 

\subsection{Combination of Policies}

\begin{figure}[h]
  \includegraphics[width=\linewidth]{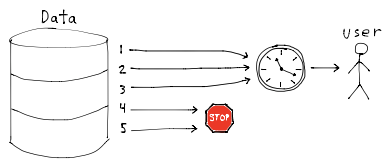}
  \caption{Sharing a Limited Number of Records Per Time Period}
  \label{fig:policy6}
\end{figure}

\paragraph{Sharing a Limited Number of Records Per Time Period}
This policy shown in Figure~\ref{fig:policy6} describes the combination of Figure~\ref{fig:policy1} and Figure~\ref{fig:policy2}.
For example, an owner of a data set would like to share a maximum of 25 records for 1 hour. 
These 25 records would be shared for that 1 hour period and then the shared records would become private.

\begin{figure}[h]
  \includegraphics[width=\linewidth]{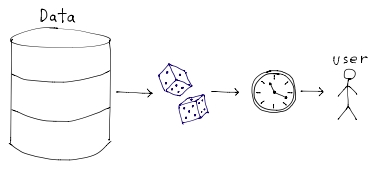}
  \caption{Sharing a Random Sample of Records Per Time Period}
  \label{fig:policy7}
\end{figure}

\paragraph{Sharing a Random Sample of Records Per Time Period}
This policy shown in Figure~\ref{fig:policy7} describes the combination of Figure~\ref{fig:policy3} and Figure~\ref{fig:policy2}.
For example, an owner of a data set would like to share 10 records randomly selected for a 1 week period.
After the 1 week period expires, the shared records would become private.

\begin{figure}[h]
  \includegraphics[width=\linewidth]{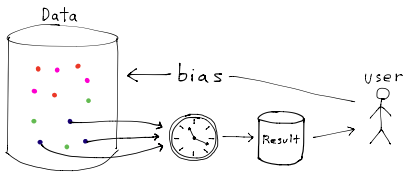}
  \caption{Sharing a Biased Sample of Records Per Time Period}
  \label{fig:policy8}
\end{figure}

\paragraph{Sharing a Biased Sample of Records Per Time Period}
This policy shown in Figure~\ref{fig:policy8} describes the combination of Figure~\ref{fig:policy4} and Figure~\ref{fig:policy2}.
For example, an owner of a data set would like to share location based data to users around the world based on their current location for a 3 day period.
After the 3 period expires, the shared records would become private.

\subsection{Complex Policies}

\begin{figure}[h]
  \includegraphics[width=\linewidth]{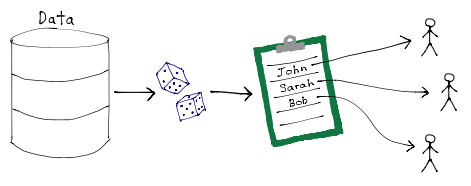}
  \caption{Sharing a Random Sample of Data with a Set of Other Users}
  \label{fig:policy9}
\end{figure}

\paragraph{Sharing a Random Sample of Data with a Set of Other Users}
This policy shown in Figure~\ref{fig:policy9} describes sharing randomly selected records to a specific set of allowed users.
For example, a social network user would like users on his/her friends list to see two random posts out of all his/her posts over the past week.

\begin{figure}[h]
  \includegraphics[width=\linewidth]{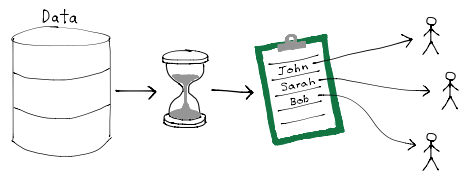}
  \caption{Sharing Data with a Set of Other Users For a Limited Time Period}
  \label{fig:policy10}
\end{figure}

\paragraph{Sharing Data with a Set of Other Users For a Limited Time Period}
This policy shown in Figure~\ref{fig:policy10} describes sharing data to a set of specific users for a limited time period.
For example, a Snapchat user would like to share his/her picture to three chosen friends for 30 seconds.

\begin{figure}[h]
  \includegraphics[width=\linewidth]{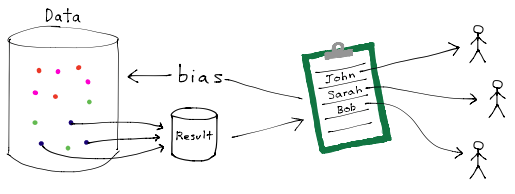}
  \caption{Sharing a Biased Sample of Data with a Set of Other Users}
  \label{fig:policy11}
\end{figure}

\paragraph{Sharing a Biased Sample of Data with a Set of Other Users}
This policy shown in Figure~\ref{fig:policy11} describes sharing data to a set of specific users based on some bias.
For example, a user of a social network would like to let users on his/her friends list know what favorite movies they have in common.
Only movies in common are shared so each friend's bias would be their own list of favorite movies.

\balancecolumns 

\end{document}